\documentclass[a4paper,12pt]{article}
\newcommand{\be}{\begin{equation}}
\newcommand{\ee}{\end{equation}}
\newcommand{\ba}{\begin{eqnarray*}}
\newcommand{\ea}{\end{eqnarray*}}

\usepackage{graphicx,pstricks}
\usepackage{setspace}
\usepackage {amssymb,amsmath}
\usepackage{parskip}
\title{ Modified holographic Ricci dark energy model and statefinder diagnosis in flat
universe.}
\author{Titus K Mathew$^1$, Jishnu Suresh$^2$ and Divya Divakaran$^3$ \\
\vspace{0.2in} \\
 Department of Physics, \\
Cochin University of Science and Technology, 
\\{\it  Kochi-22, India}.\\
\vspace{0.1in} \\
E-mail: $^1$titus@cusat.ac.in, tituskmathew@gmail.com, \\
$^2$jishnusuresh.vasthavya@gmail.com, \\
$^3$divdnair@gmail.com.}
\date{}
\begin{document}

\maketitle

\begin{abstract}
 Evolution of the universe with modified holographic Ricci dark energy model is 
 considered. Dependency of the equation of state parameter and deceleration parameter 
 on the redshift and model parameters are obtained. It is shown that the density 
 evolution of both the non-relativistic matter 
 and dark energy are same until recent times. The evolutionary trajectories  of the 
 model for different model parameters are obtained 
 in the statefinder planes, $r-s$ and $r-q$ planes. The present statefinder parameters 
 are obtained for different model parameter values,
 using that the model is differentiated from other standard models like $\Lambda$CDM 
 model etc. We have also shown that the evolutionary trajectories are depending on the model
 parameters, and at past times the dark energy is 
 behaving like cold dark matter, with equation of state equal to zero. 
 
 \vspace{0.2in}
 
 
 \vspace{0.15in}
 
\end{abstract}

\thispagestyle{empty}

\newpage

\section{Introduction}

 Observations of distant type Ia supernovae (SNIa) and cosmic microwave 
 background anisotropy have shown that the present universe is accelerating
 \cite{Perl98}.
 This expansion may be driven by a component with negative pressure, called dark energy. 
 The simplest model of dark energy is the cosmological constant $\Lambda$ which can fit 
 the observations 
 in a fair way \cite{wein1, sahni1}, whose equation of state is $\omega_{\Lambda} =
 -1.$
 during the evolution of the universe.
 However there are two serious problems with 
 cosmological constant model, namely the fine tuning and the cosmic coincidence 
 \cite{Cope1}.
 To solve these problems different dynamic dark energy models have been proposed, with 
 varying equation of state during the expansion of the universe. Holographic dark 
 energy (HDE) is one among them \cite{Cohen1,Hsu1, Li1}. HDE is constructed based on the 
 holographic principle, that in quantum gravity, the entropy of a system scales not with
 its volume but with its surface area $L^2$, analogically the cosmological constant in 
 Einstein's theory also is inverse of some length squared. It was shown that
 \cite{Cohen1}
 in effective 
 quantum field theory, the zero point energy of the system with size $L$ should not 
 exceed the mass of a black hole with the same size, thus $L^3 \rho_{\Lambda} 
 \leq L M_P^2,$ where $\rho_{\Lambda}$ is the quantum zero-point energy and 
 $M_P = 1/ \sqrt{8\pi G}$, is the reduced Plank mass. This inequality relation implies a 
 link between the ultraviolet (UV) 
 cut-off, defined through $\rho_{\Lambda}$ and the infrared (IR) cut-off encoded in
 the scale 
 $L.$ In the context of cosmology one can take the dark energy density of the universe
 $\rho_X$ as the same as the vacuum energy, i.e. $\rho_x = \rho_{\Lambda}.$ The largest
 IR cut-off $L$ is chosen by saturating the inequality, so that the holographic energy
 density can be written as 
 \begin{equation}
  \rho_x = 3c^2M_P^2L^{-2}
 \end{equation}
where $c$ is numerical constant. In the current literature, the IR cut-off has been 
taken as the Hubble horizon \cite{Hsu1,Li1}, particle horizon and event horizon 
\cite{Li1} or some generalized IR cut off \cite{Gao1,Linsen1,Yang1}. The HDE models 
with 
Hubble horizon or particle horizon as the IR cut-off, cannot lead to the current
accelerated expansion \cite{Hsu1} of the universe. When the event horizon is taken as 
the length scale, the model is suffered from the following disadvantage. Future event 
horizon is a global concept of 
space-time. On the other hand density of dark energy is a local quantity. So the 
relation between them will pose challenges to the concept of causality. These leads to
the introduction new HDE, where the length scale is given by the average radius of the
Ricci scalar curvature, $R^{-1/2}.$

The holographic Ricci dark energy model introduced by Granda and Oliveros \cite{Granda1}
based on the space-time scalar curvature, is fairly good in fitting with the 
observational 
data. This model have the following advantages. First, the fine tuning problem can be 
avoided in this model. Moreover, the presence of event horizon is not presumed in this 
model, so that the causality problem can be avoided. The coincidence problem can also 
be solved effectively in this model. Recently a modified form of Ricci dark energy was 
studied \cite{Chimento1} in connection with the dark matter interaction, and analyses 
the model using \emph{Om} diagnostic. In this paper
we have considered the evolution of the universe in Modified Holographic Ricci Dark 
Energy (MHRDE) model and 
obtain the statefinder parameters to discriminate this model with other standard dark 
energy models. 

Statefinder parameters is a sensitive and diagnostic tool used to discriminate various 
dark energy models. The Hubble parameter $H$ and deceleration parameter $q$ alone 
cannot discriminate various dark energy models because of the degeneracy on these 
parameters. Hence Sahni et al. \cite{Sahni2} introduces a set of parameters $\{r,s\}$
called statefinder parameters, defined as,
\begin{equation} \label{statefinder}
 r = { \dddot a \over a H^3}, \, \, \, \, \, \, \, \, \, \, \, \, \, \, \, \, 
 s = {r - \Omega_{total} \over 3 (q - 
 \Omega_{total} ) /2 },
\end{equation}
where $a$ is the scale factor of the expanding universe and $\Omega_{total}$ is the 
total energy density containing dark energy, energy corresponds to curvature and also 
matter (we are neglecting the radiation part in our analysis).
In general statefinder parameter is a geometrical diagnostic such that it depends upon 
the expansion factor and hence on the metric describing space-time. The $r-s$ plot of 
dark energy models can help to differentiate and discriminate various models. 
For the well known $\Lambda$CDM model, the $r-s$ trajectory
is corresponds to fixed point, with $r=1$ and $s=0$ \cite{Sahni2}. The cosmological 
behavior of various dark models including holographic dark energy model, were  
studied and differentiated in the recent literature using statefinder parameters 
\cite{Setare1,Malekjani1,Huang1,
Mub1}.

The paper is organized as follows. In
section 2, we have studied the cosmological behavior of the MHRDE model and in 
section 3  
we have considered the statefinder diagnostic analysis followed by the conclusions
in section 4.

\section{The MHRDE model}

The universe is described by the Friedmann-Robertson-Walker metric given by
\begin{equation}
 ds^2 = -dt^2 + a(t)^2 \left( {dr^2 \over 1-kr^2} + r^2 d\theta^2 + r^2 \sin^2\theta 
 d\phi^2 \right),
\end{equation}
where $(r, \theta, \phi)$ are the co-moving coordinates, $k$ is the curvature 
parameter with 
values, $k=1,0,-1$ for closed, flat and open universes respectively and $a(t)$ is the 
scale factor, with $a_0 = 1,$ is taken as its present value. The Friedmann equation 
describing the evolution of the universe is
\begin{equation}
 H^2 + {k \over a^2} = {1 \over 3} \sum_i \rho_i,
\end{equation}
where we have taken $8\pi G =1$, the summation includes the energy densities of 
non-relativistic matter and
dark energy, i.e. $\sum_i \rho_i = \rho_m + \rho_x$ . The modified 
holographic Ricci dark energy can be expressed by taking the IR cutoff with the modified 
Ricci
radius in terms of $\dot{H}$ and $H^2$ as \cite{Granda1,Chimento1}
\begin{equation}
 \rho_x = {2 \over \alpha - \beta } \left( \dot{H} + {3 \alpha \over 2} H^2 \right),
\end{equation}
where $\dot{H}$ is the time derivative of the Hubble parameter, $\alpha$ and $\beta$
are free constants, the model parameters. Chimento et. al. studied this type of dark 
energy in interaction 
with the dark matter with Chaplygin gas \cite{Chimento1}, but our analysis is mainly 
concentrated on the cosmological evolution of MHRDE and analyses with statefinder
diagnostic. Substituting the dark energy density as the MHRDE in the Friedmann equation,
and changing the variable form cosmic time $t$ to $x=\ln a$ 
we get
\begin{equation}
 H^2 + {k \over a^2} = { \rho_m \over 3} + {1 \over 3(\alpha - \beta)} {dH^2 \over dx} + { \alpha \over
 \alpha - \beta} H^2
\end{equation}
Introducing the normalized Hubble parameter as $ h = H/H_0 $ and $\Omega_k = -k/H_0^2$
, where $H_0$ is the Hubble parameter for $x=0$, the above equation become,
\begin{equation} \label{dif1}
 h^2 - \Omega_{k0} e^{-2x} = \Omega_{m0} e^{-3x} +  {1 \over 3(\alpha - \beta)} 
 {dh^2 \over dx} + { \alpha \over
 \alpha - \beta} h^2,
\end{equation}
where $\Omega_{mo} = \rho_{m0}/3H_0^2$ is the current density parameter of 
non-relativistic matter ( we will take 0.27 as its values for our analysis throughout.)
with current density $\rho_{m0}$  and $\Omega_{ko}$ 
is the present relative density parameter of the curvature. We will consider only flat 
universe, where $\Omega_{k0}=0$ in our further analysis.
Solving the first order differential equation (\ref{dif1}) we obtain the dimensionless
Hubble parameter $h$ as,
\begin{equation} \label{hsquare}
 h^2 = \Omega_{mo} e^{-3x} + { \alpha - 1 \over 1 - \beta} 
 \Omega_{m0} e^{-3x} + \left[ {(\alpha - \beta) \Omega_{m0} \over \beta - 1 } + 1 
 \right] e^{-3\beta x}.
\end{equation}
Comparing this with the standard Friedman equation, the dark energy density can be 
identified as
\begin{equation} \label{darke1}
 \Omega_x = { \alpha - 1 \over 1 - \beta } \Omega_{m0} e^{-3x} + \left[ { (\alpha - 
 \beta)
 \Omega_{m0} \over \beta - 1} + 1 \right] e^{-3\beta x}
\end{equation}
This shows that similar to the result obtained in references \cite{Gao1,Malekjani1} 
for Ricci dark energy, the 
MHRDE density has one part which evolves like non-relativistic matter 
($ \sim e^{-3x}$) and the 
other part is slowly increasing with the decrease in redshift. The pressure 
corresponding the dark energy can be calculated as,
\begin{equation}
 p_x = - \Omega_x - \frac{1}{3} {d\Omega_x \over dx} 
     = \left[ (\alpha - \beta ) \Omega_{m0} + \beta -1 \right] e^{-3\beta x}
\end{equation}
Form the conservation equation, we can obtain the corresponding equation 
of state parameter for the flat universe, using equation (\ref{darke1}) as,
\begin{equation} \label{eos1}
 \omega_x = -1 - {1 \over 3} {d \ln \Omega_x \over dx} 
          = -1 + \left\{ { ( \alpha - 1 ) \Omega_{m0} + \beta 
          \left[ (1 - \beta) - ( \alpha - \beta) \Omega_{m0} \right] e^{3(1 - \beta )x} 
          \over 
          (\alpha - 1) \Omega_{m0} +  \left[ (1-\beta ) - ( \alpha
          - \beta) \Omega_{m0} \right] e^{3(1 - \beta )x} } \right\}
\end{equation}
This equation of state implies the possibility of transit form $\omega_x > -1 $  
 to 
$\omega_x < -1,$ corresponds to the phantom model \cite{Wang1,Wang2} for suitable 
model parameter values. Recent observational 
evidences shows that the dark energy equation of state parameter can crosses the 
value -1 
\cite{Alam1}. In a universe dominated with MHRDE, where the contribution from
the non-relativistic matter behavior term is negligible in the dark energy density, 
the equation state parameter 
become, 
\begin{equation} \label{reducedeos}
 \omega_x = -1 + \beta
\end{equation}
So if $\beta$ is less than zero, the equation of state can crosses the phantom divide.
In the far future of the universe, when redshift $z \rightarrow -1$ also, the 
equation of state parameter reduces to the form given in equation (\ref{reducedeos}). 
So the behavior of the dark energy is depending strongly on the model parameter 
$\beta.$

\begin{figure}[here]
 \includegraphics[scale=0.8]{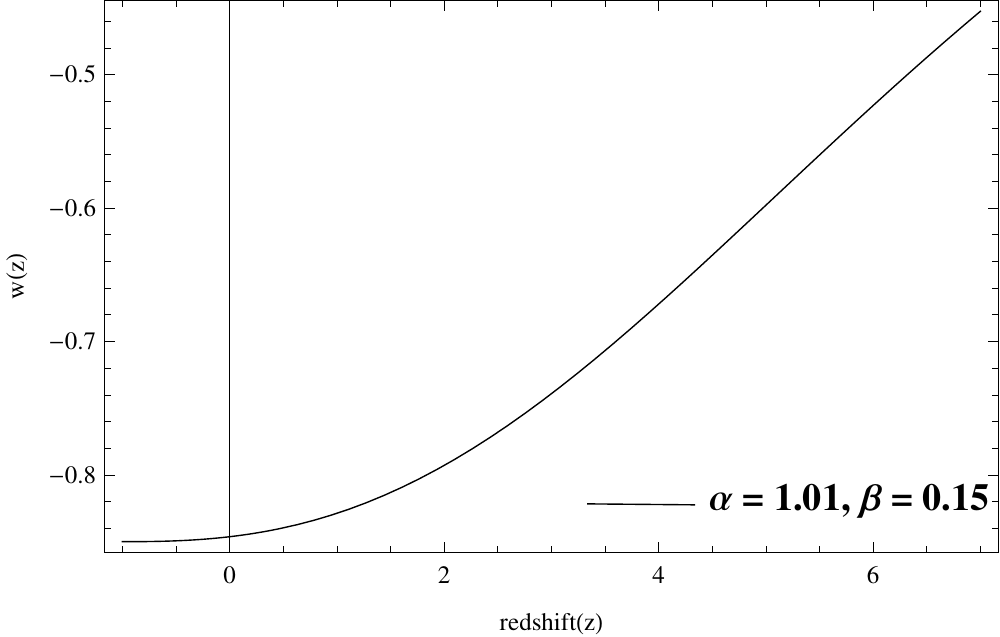}
 \caption{ \emph{Evolution of equation of state parameter $\omega_x$ with redshift $z$ 
 for the best fit values $\alpha = 1.01$ and $\beta = 0.15.$}}
 \label{fig:eos1}
\end{figure}
We have plotted the evolution of the equation of state parameter of MHRDE 
with redshift in figure \ref{fig:eos1},
using the 
best fit values of the model parameters $\alpha$ and $\beta$ as \cite{Chimento1} 
$(\alpha, \beta ) = 
(1.01, 0.15)$  and $\Omega_{m0}$=0.27.
The evolution of $\omega_x$ of dark energy shows that in the remote past of the 
universe, that is at large redshift, the equation of 
state parameter is near zero, implies that the dark energy behaves like the 
cold dark matter in the remote past.
The plot also shows that at far future of the universe as $z \rightarrow -1,$ 
the equation of state parameter approaches a saturation value. The present value of 
the equation of state parameter according to this plot is negative, and is around
$\omega_x = -0.7.$  \\
\begin{figure}[here]
 \includegraphics[scale=0.8]{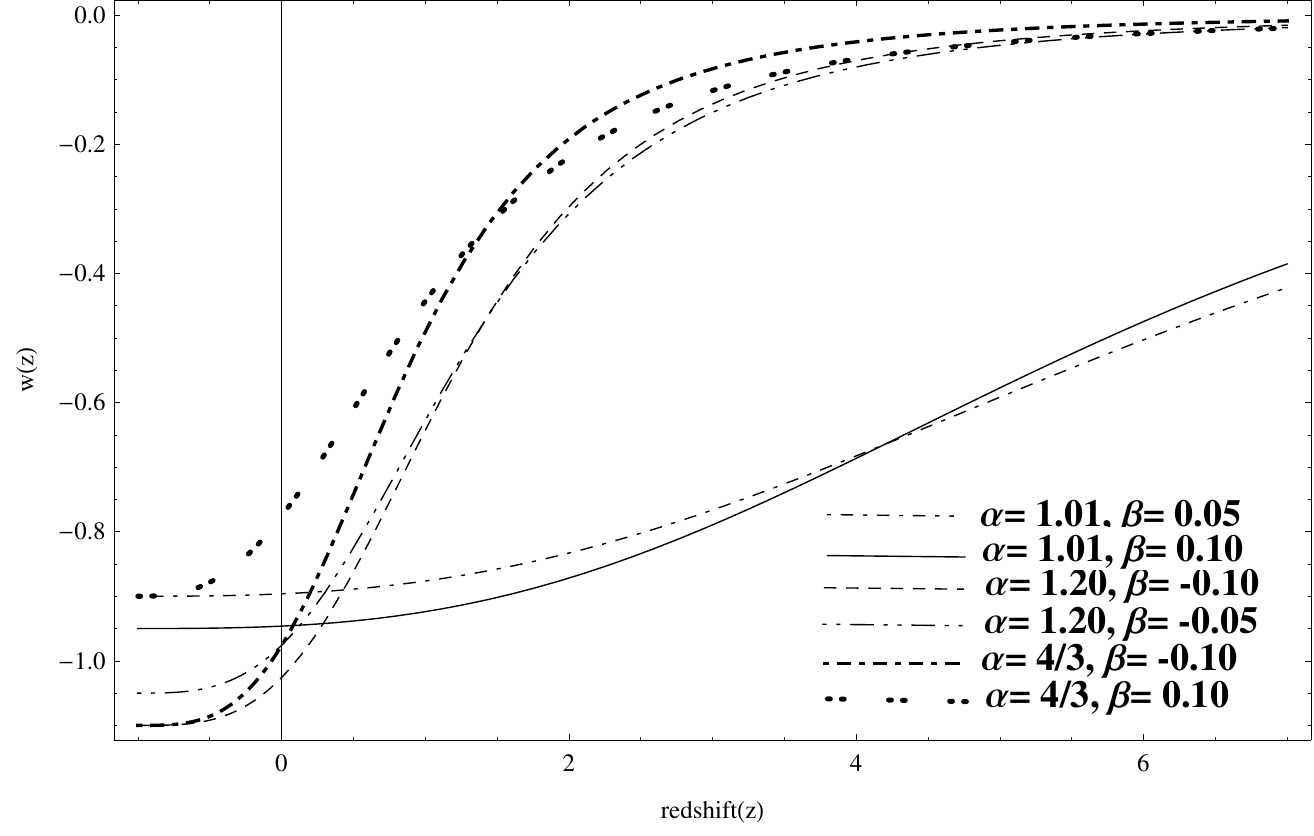}
 \caption{\emph{Evolution of the equation of state parameter for other values of 
 $\alpha$ and 
 $\beta$}}
 \label{fig:eos}
\end{figure}

For other values of the model parameters $(\alpha, \beta)$ \cite{Chimento1} as 
($\alpha, \beta)$
= (1.01, 0.05), (1.01,0.1), (1.2,-0.05), (1.2,-0.1), (4/3,0.1) and (4/3, -0.1) 
the behavior of
the equation of state parameter is given figure \ref{fig:eos}. For a given value of
$\alpha$ 
the saturation value of $\omega_x$ in the future universe decreases as $| \beta 
|$ increases.

Figure \ref{fig:eos} shows that irrespective of 
the values of the parameters $(\alpha, \beta),$ the equation of state parameter is 
negative at present times implies that the present universe is accelerating, and 
also in the remote 
past at high 
redshift $\omega_x \rightarrow 0,$ indicate that MHRDE behaves like cold dark matter
in the past stages of the universe. 
For negative values of $\beta$ the 
equation of state parameter  crosses -1, in that case
 it can be classified as quintom \cite{Feng1} dark energy and 
for the case $\omega_x < -1$ the universe will evolve into 
a phantom energy dominated epoch \cite{Caldwel1}.

\begin{figure}
 \includegraphics[scale=0.9]{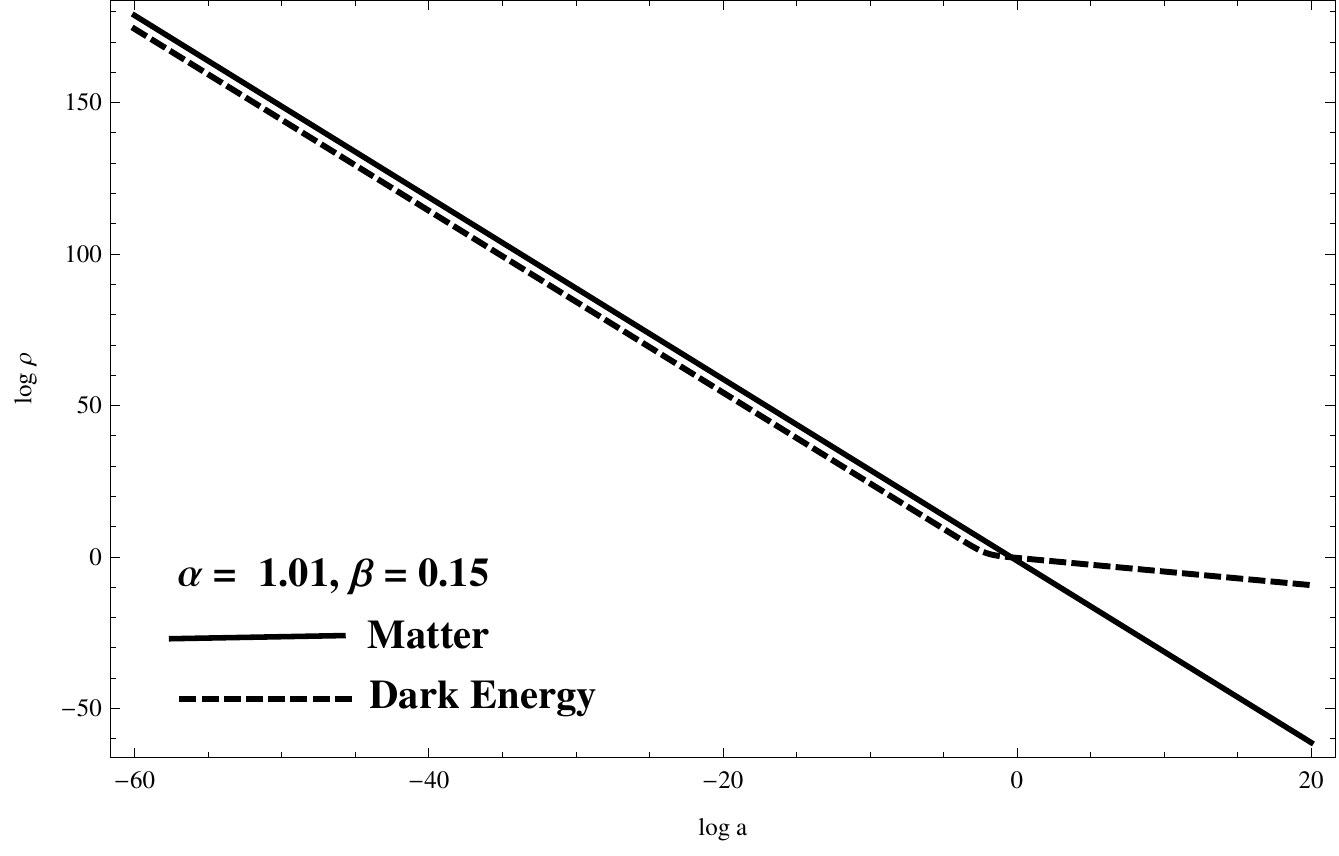}
 \caption{\emph{Evolution of non-relativistic matter density and MHRDE density in
 log.scale}}
 \label{fig:densityevolve}
\end{figure}
In figure \ref{fig:densityevolve}, we have shown a comparison of the evolution of 
non-relativistic matter
density and MHRDE density in logarithmic scale. Here we have neglected phase transitions, 
transitions from non-relativistic to relativistic particles at high temperatures and 
new degrees of freedom etc. It is expected that these would not make much qualitative 
difference in the result. The plot shows that in the present model the densities of 
non-relativistic matter and dark energy were comparable with each other in the past 
universe that is at high redshift. The acceleration began at low redshifts, which solves
the coincidence problem.

The deceleration parameter $q$ for the MHRDE model can obtained using the relation
\begin{equation}
 q = -{ \dot{H} \over H^2} - 1.
\end{equation}
This equation can be expressed in terms of the dimensionless Hubble parameter $h$ as 
\begin{equation} \label{qpara}
 q = - {1 \over 2h^2} {dh^2 \over dx} - 1
\end{equation}
Using equation (\ref{hsquare} ) the above equation can be written as 
\begin{equation}
 q = {  \left( { \alpha - \beta \over 1 - \beta} \right) \Omega_{m0}
       e^{-3x} + \left[ { (\alpha - \beta ) \Omega_{m0} \over \beta - 1)} + 1 
       \right] (3\beta - 2) e^{-3\beta x}
       \over
       2 \left [ \left( { \alpha - \beta \over 1 - \beta} \right) \Omega_{m0} e^{-3x}
       + \left( { (\alpha - \beta ) \Omega_{m0} \over \beta -1 } + 1 \right ) 
       e^{-3\beta x} \right] }
\end{equation}
This equation shows the dependence of the deceleration parameter on the model 
parameters $\alpha$ and $\beta.$ As an approximation, if we neglect the contribution
form the first terms in both numerator and denominator (since they are negligibly small)
the deceleration parameter will become $\displaystyle q = (3\beta - 2)/2.$ Which shows,
as
$\beta$ increases form from zero, the parameter $q$ increases form -1, that is the 
universe enter the acceleration phase at successively later times. 
\begin{figure}[here]
 \includegraphics[scale=0.8]{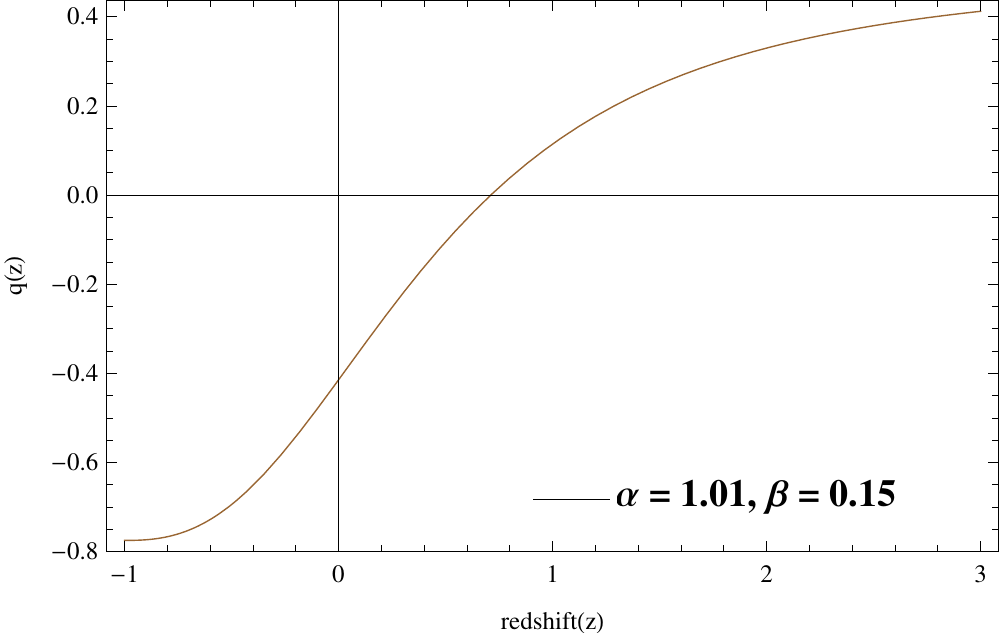}
 \caption{\emph {Evolution of deceleration parameter $q$ for the best fit model 
 parameters $\alpha$=1.01 and $\beta$=0.15}}
 \label{fig:deceleration1}
\end{figure}
\begin{figure}[here]
 \includegraphics[scale=0.8]{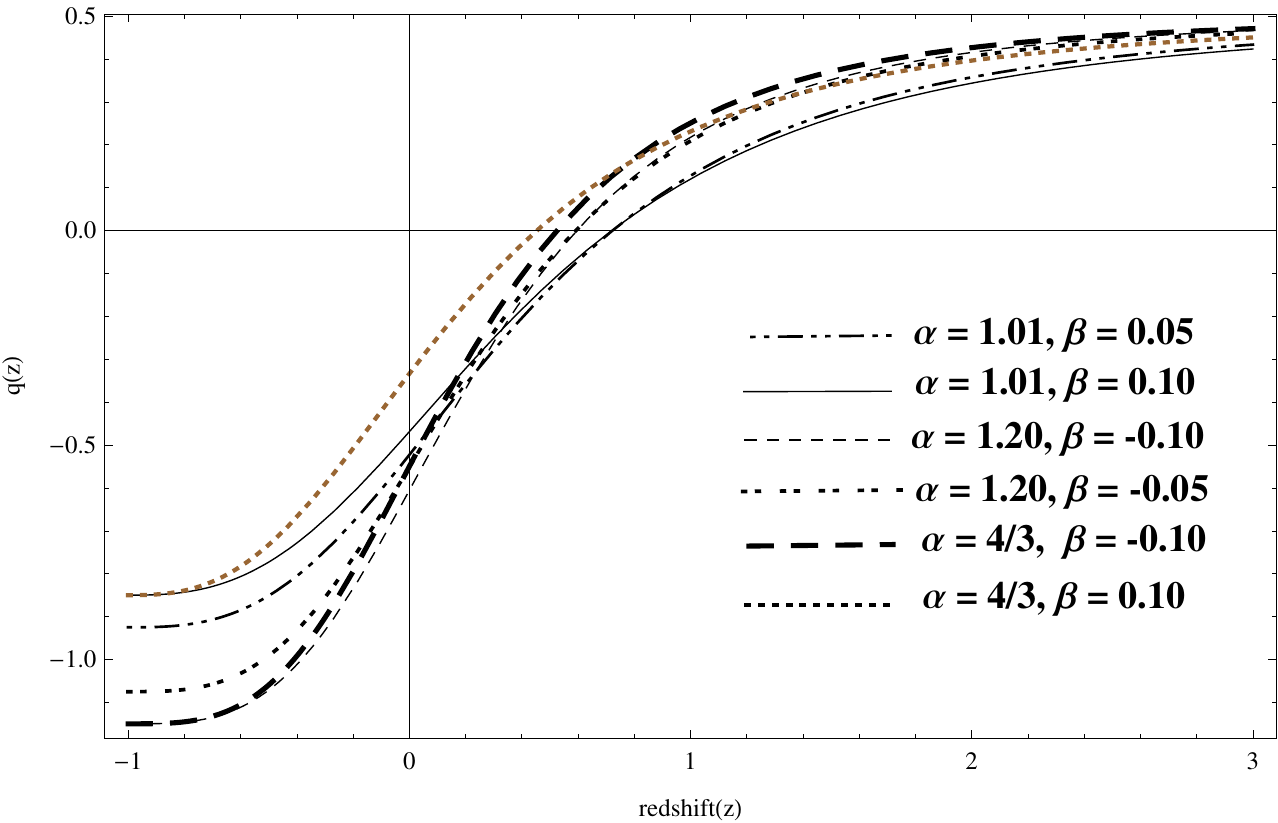}
 \caption{\emph{Evolution of deceleration parameter for other values of the model
 parameter.}}
 \label{fig:deceleration}
\end{figure}
In figure 
\ref{fig:deceleration1} and figure \ref{fig:deceleration} we
have plotted the evolution of the deceleration
parameter with redshift. Figure \ref{fig:deceleration1} is for the best fitting model
parameters 
$\alpha$=1.01, $\beta$=0.15 and figure \ref{fig:deceleration} is for the 
remaining model parameter values. The plots shows that at large redshift, 
the deceleration
parameter approaches 0.5. The universe is entering the acceleration in the recent past
at $z<1.$ The plot also shows that as $\alpha$ increase, the entry to the accelerating 
phase is occurring at relatively lower values of redshift, that is the universe entering 
the accelerating phase at relatively later times as the parameter $\alpha$ increases. 
The transition of the universe 
from deceleration to the accelerating phase is occurred at the the redshift $Z_T = 0.76$,
for the best fit model parameters . For comparison 
the combined analysis of SNe+CMB data with 
$\Lambda$CDM model gives the range $Z_T(\Lambda CDM)$ = 0.50 - 0.73 \cite{Alam1, Zhu12}. For taking
consideration of the entire model parametric range, the transition to the accelerating 
phase can be obtained, as in figure \ref{fig:deceleration} as $Z_T (MHRDE) $=0.50 - 0.76.
The comparison of the two ranges shows that in the MHRDE model the universe entering 
the accelerating 
expansion phase earlier than in the $\Lambda$CDM model.
The present value of the 
deceleration parameter for the best fit model parameters $\alpha$=1.01, $\beta$=0.15 is
$q_0 = -0.45$ as from figure \ref{fig:deceleration1}.

\section{Statefinder diagnostic}

We have calculated the statefinder parameters $r$ and $s$, as defined earlier in 
equation (\ref{statefinder}). 
Statefinder parameters can provide us with a diagnosis which should unambiguously probe
the properties of various classes of dark energy models. Equation (\ref{statefinder}) 
for
$r$ and $s$ can be 
rewrite in terms of $h^2$ as,
\begin{equation} 
 r  =  {1 \over 2 h^2} {d^2 h^2 \over dx^2} + {3 \over 2 h^2} {d h^2 \over dx} +1  
 \end{equation}
and
\begin{equation}
 s  =  - \left\{ {  {1 \over 2 h^2} {dh^2 \over dx^2}+ {3 \over h^2} {dh^2 \over dx} 
        \over
        {3 \over 2h^2} {dh^2 \over dx} + {9 \over 2}  } \right\} 
\end{equation}
On substituting the relation for $h^2$ from equation (\ref{hsquare}), the above 
equations for a flat universe (in which $\Omega_k = 0$) become
\begin{equation} \label{r}
 r = 1 + \left\{ 
 { 9 \beta (\beta -1) \left( { (\alpha - \beta) \Omega_{m0} \over
     \beta - 1} + 1 \right) e^{-3\beta x}
     \over
     2 \left[ \Omega_{m0} e^{-3x} + \left( { \alpha - 1 \over 1- \beta} \right) 
     \Omega_{m0} e^{-3 \beta x} + \left( {( \alpha - \beta) \Omega_{m0} \over \beta - 1 }
     + 1 \right) 
     e^{-3 \beta x}  \right] }
     \right\}    
\end{equation}
and $s$ is become,
\begin{equation} \label{s}
 s = - \left\{ {  \beta (\beta - 1) \left( { (\alpha - \beta ) \Omega_{m0} \over 
      \beta - 1 } + 1 \right) e^{-3\beta x}   
      \over
       \left[ { (\alpha - \beta ) \Omega_{m0} \over \beta -1 } + 1 \right] (1-\beta)
      e^{-3\beta x}  } \right \} = \beta
\end{equation}
From equations (\ref{r}) and (\ref{s}), it is evident that $r=1, \, \, s=0$ if $\beta
= 0$ 
and no matter what value $\alpha$ is, and this point in the $r-s$ plane is corresponds
to the
$\Lambda$CDM model. This point is a very fixed point, thus statefinder diagnostic fails 
to discriminate between $\Lambda$CDM model and MHRDE model for the model parameter value
$\beta=0.$ Since $s$ is a constant for flat universe in this model, the trajectory in
the $r-s$ plane is a vertical segment, with constant $s$ during the 
evolution of the universe, while $r$ is monotonically 
decreasing form 1, if $\beta$ is positive and monotonically increasing if $\beta$ is
assuming negative values. For a simple understanding, let us assume that $\Omega_m$ 
contribution is
negligible small, when dark energy is dominating, then the equation (\ref{r}) reduces 
to 
\begin{equation}
 r = 1 + {9 \beta (\beta -1) \over 2 }
\end{equation}
In this case for $\beta$=0.05, 0.1 and 0.15 the 
corresponding values of $r$ are 0.79, 0.60, 0.43 respectively. Bur when $\beta$ 
assumes the negative values -0.05 and -0.10, the corresponding values of $r$ become
1.24 and 1.5 respectively. So at the outset the MHRDE model gives a $r-s$ trajectory, 
as $r$ starting 
form 1 and due to evolution of the universe the $r$ will  decreases to 
$1 - {9 \beta (1-\beta ) \over 2 }$ if $\beta$ is positive and  increases to
$1 + {9 \beta (\beta -1) \over 2 },$ if $\beta$ is negative.

\begin{figure}
 \includegraphics[scale=0.8]{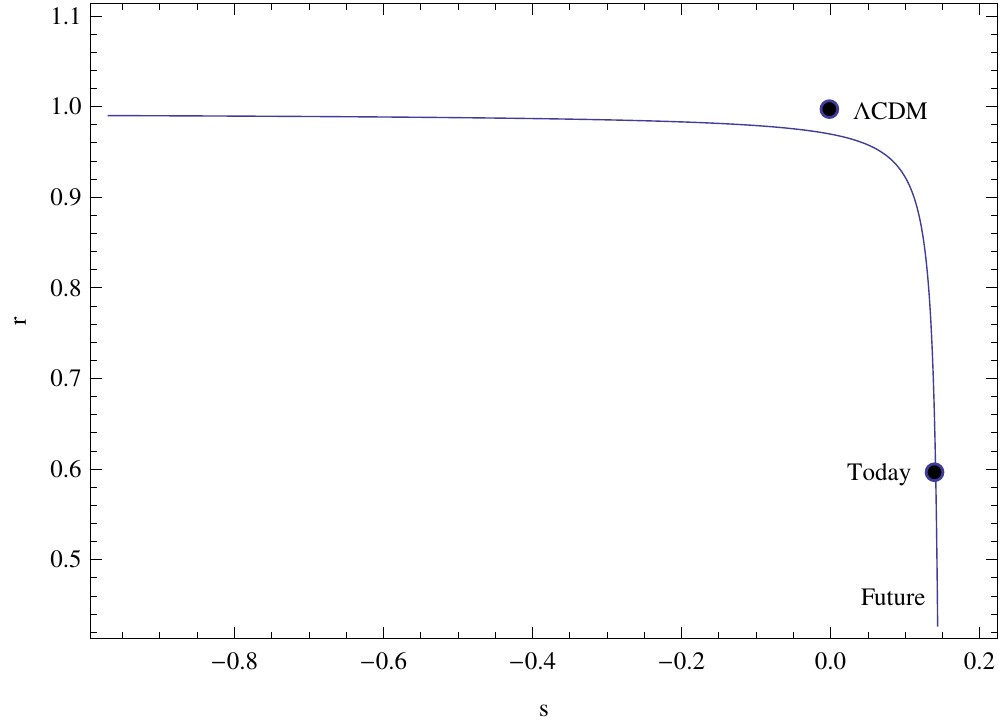}
 \caption{\emph{Evolutionary trajectory in the $r-s$ plane for MHRDE model for the best
 fit values of the model parameters $\alpha$=1.01, $\beta$=0.15.
 The black spot on the top right
 corner corresponds to $r=1, \, s=0$ the $\Lambda$CDM model. The today's point 
 corresponds to $r$=0.59, $s$=0.15}}
 \label{fig:rs1}
\end{figure}

The  $r-s$ evolutionary trajectory in the MHRDE model in flat universe for 
the best fit model 
parameters $\alpha$=1.01 and $\beta$=0.15, is given in figure \ref{fig:rs1}. In this 
plot as the universe expands, the trajectory in the $r-s$ plane starts form left 
to right. The standard $\Lambda$CDM model is corresponds to $r=1, \, s=0$ is denoted.
In this model the parameter $r$ first decrease very slowly with $s$, then after 
around $s$=0 $r$ decreases steeply. The today's value of the 
statefinder parameter ($r_0$=0.59, $s_0$=0.15) is denoted in the plot.

\begin{figure}[here]
\centering
\includegraphics[scale=0.8]{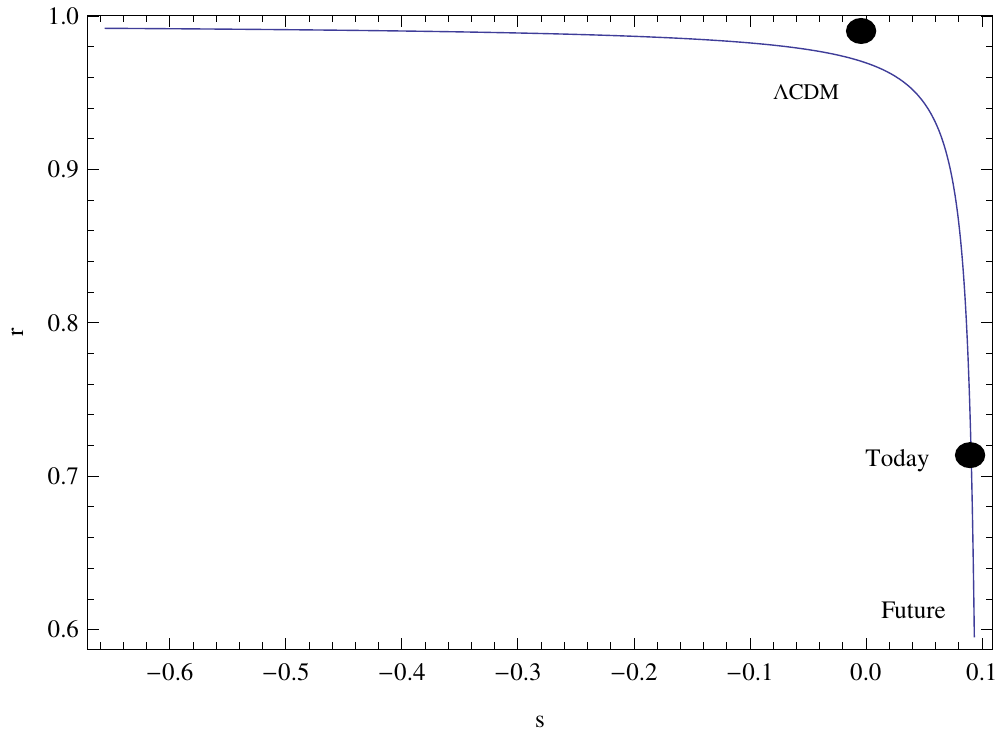}
\includegraphics[scale=0.8]{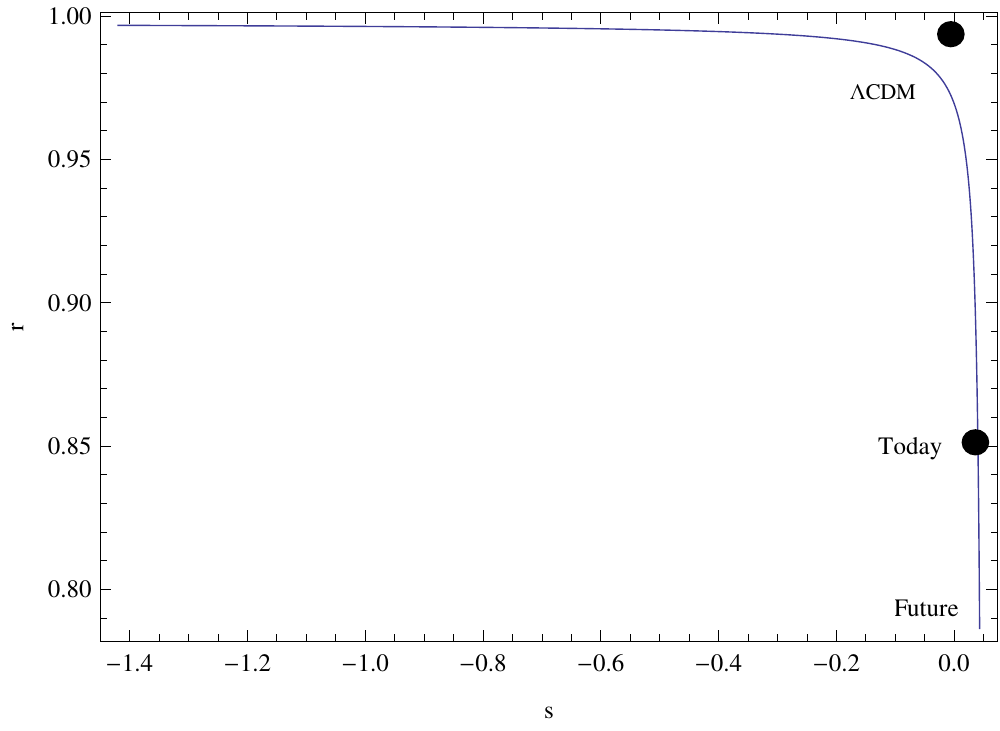}
\caption{\emph{the first plot is for $\alpha, \beta = 1.01, 0.10$ and the second plot is for
$\alpha, \beta = 1.01, 0.05$ The black spot on the top right corner corresponds to 
$\Lambda$CDM model, the present state of the evolution is denoted as today's point.}}
\label{fig:rs2}
\end{figure}

For other model parameters, the $r-s$ plots are given in figure \ref{fig:rs2}. These 
plots also 
shows the same behavior of figure \ref{fig:rs1}, but the separation between $\Lambda$CDM 
model and MHRDE model in the $r-s$ plane is increasing as $\beta$ increases. The 
respective todays universe corresponds 
$r_0,s_0 = 0.71,0.1$ and $r_0,s_0 = 0.85,0.05.$ 

\begin{figure}
 \centering
 \includegraphics[scale=0.8]{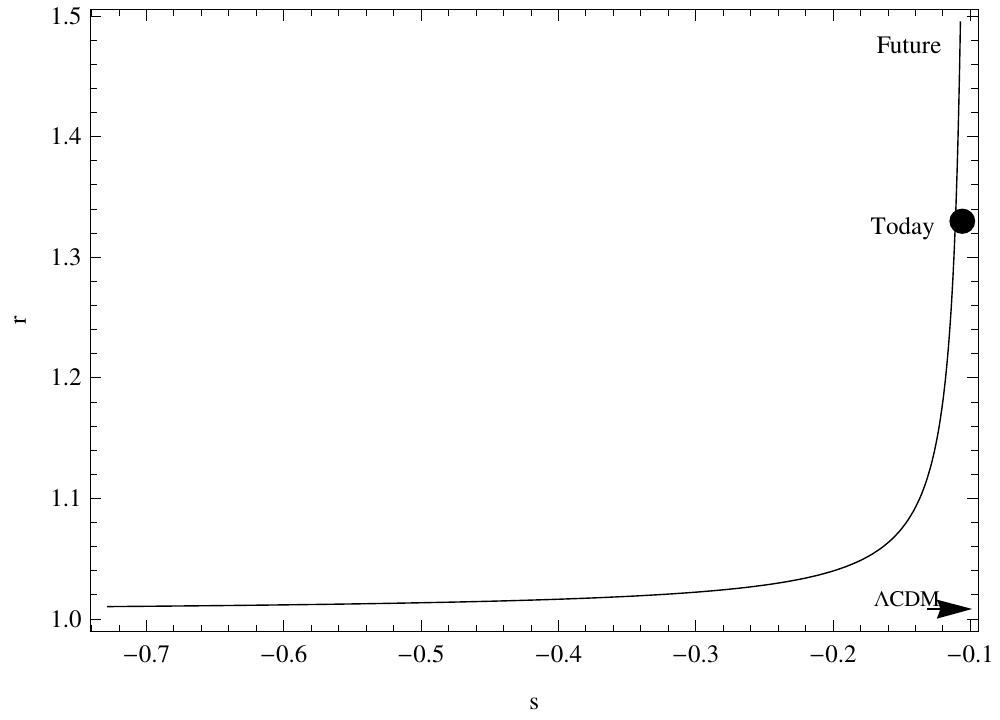}
 \includegraphics[scale=0.8]{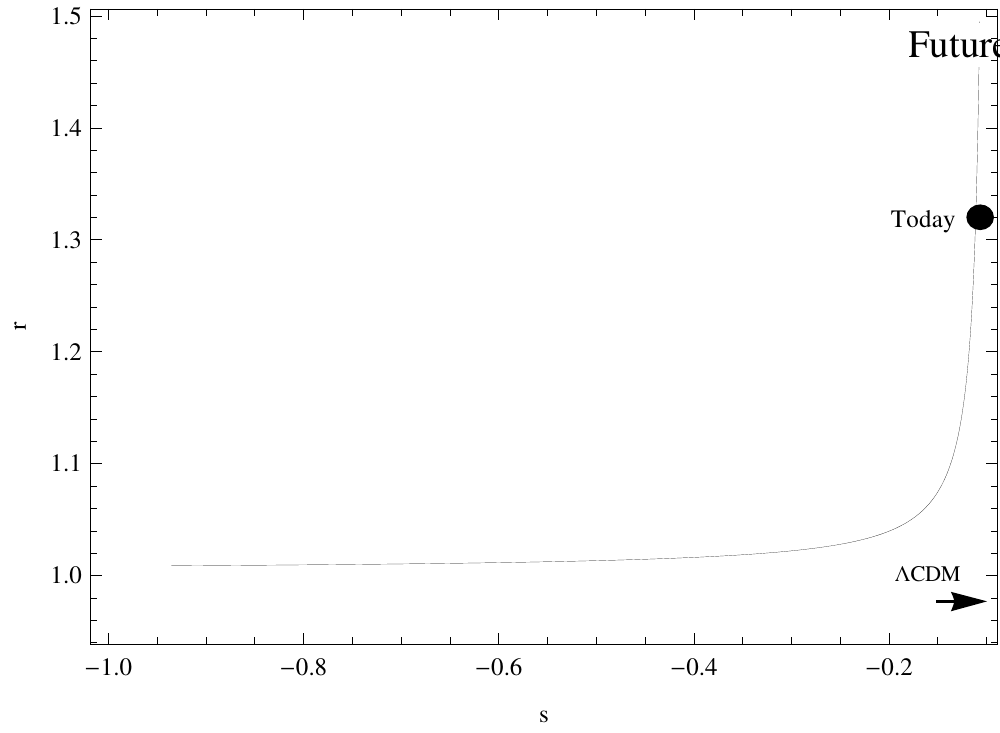}
 \caption{\emph{$r-s$ plots for model parameters $\alpha, \beta = 1.2, -0.10; \, 
 4/3, -0.10.$ 
 the arrow in lower left corner of the panel shows the evolution towards $\Lambda$CDM 
 model. The present position of the evolution is denoted as today's point.}}
 \label{fig:rs3}
\end{figure}

For negative values of $\beta$, the evolutionary characteristics is plotted
in figure \ref{fig:rs3} for model parameters $\alpha, \beta = 1,2, -0.10 ; 4/3, -0.10.$ 
Here also the evolution in the $r-s$ plane is from left to right. In this case the
behavior
is different form that for the positive $\beta$ value, in the sense that as $s$ 
increases The $r$ is increasing to vales greater than one.
The increase is very slowly at first then increases steeply as the universe evolves.
The today's value in these cases are $r_0$=1.325, $s_0$=-0.10 when ($\alpha, \beta$)=
(1.2,-0.10) and $r_0$=1.321, $s_0$=-0.10 for ($\alpha, \beta$)=(4/3, -0.10) respectively.
The difference between 
MHRDE model for these model parameters and $\Lambda$CDM can be noted. 

The statefinder diagnostic can discriminate this model with other models. As 
example, for the quintessence model the $r-s$ trajectory is lying in the region
$s>0, r<1$ and for Chaplygin gas the trajectory is in the region $s<0, r>1.$ Holographic
dark energy with the future event horizon as IR cutoff, starts its evolution 
form $s=2/3, r=1$ and ends on at $\Lambda$CDM model fixed point in the future 
\cite{Huang1,Liu2}.

\begin{figure}
\includegraphics[scale=0.8]{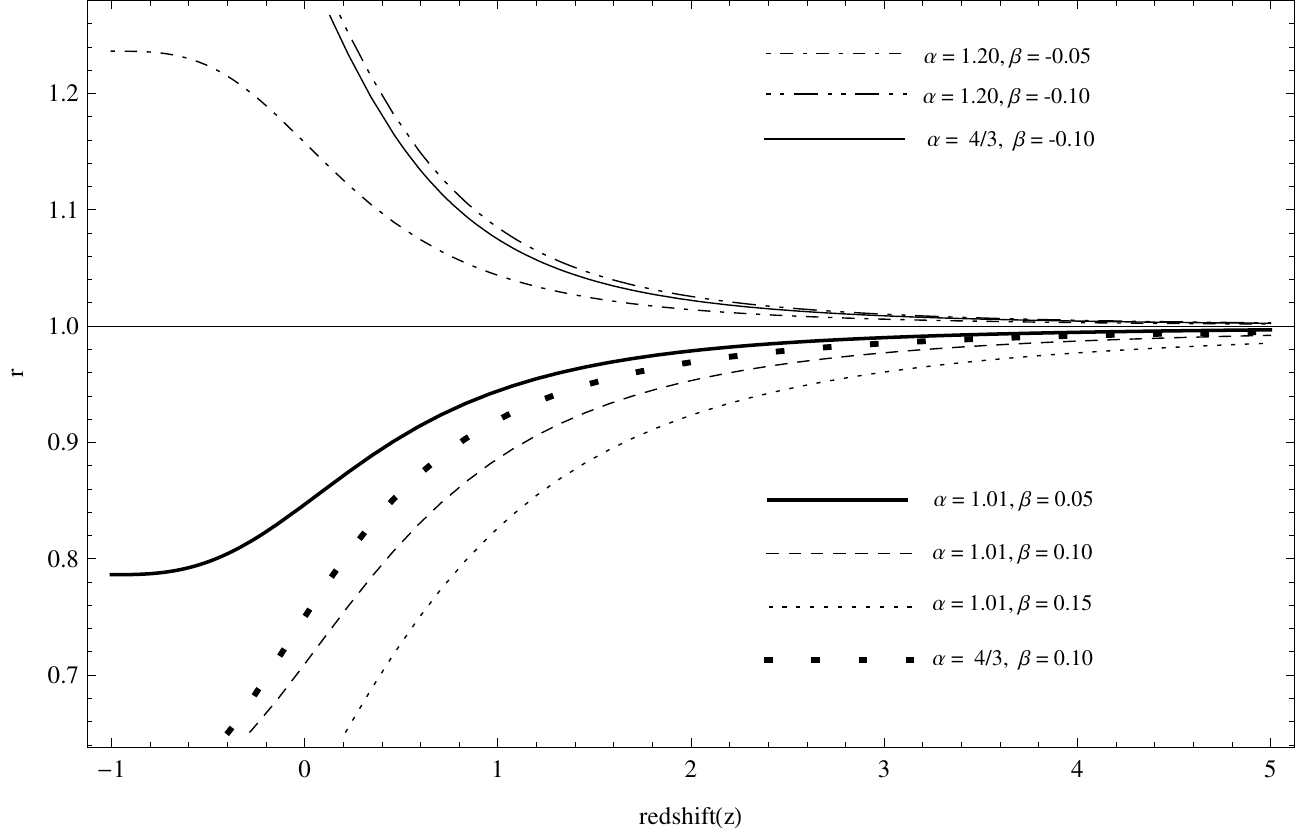}
 \caption{\emph{$r-z$ plot, all model parameters. Shows that for positive values of 
 $\beta$
 $r$ is decreasing form 1, but for negative values of $\beta$ the value of $r$ is 
increases form 1.}}
 \label{fig:rz}
\end{figure}

In order to confirm the $r-s$ behavior of MHRDE model, we have plotted the 
behavior in $r-z$ plane, 
in figure \ref{fig:rz}. For MHRDE model, 
 the $r$ value is commencing from 1 irrespective of the values of $\alpha$ and
$\beta$ at remote past and  as the universe evolves, $r$ is decreasing, if 
$\beta$ is positive and increasing, if $\beta$ is negative.

\begin{figure}
 \includegraphics[scale=0.8]{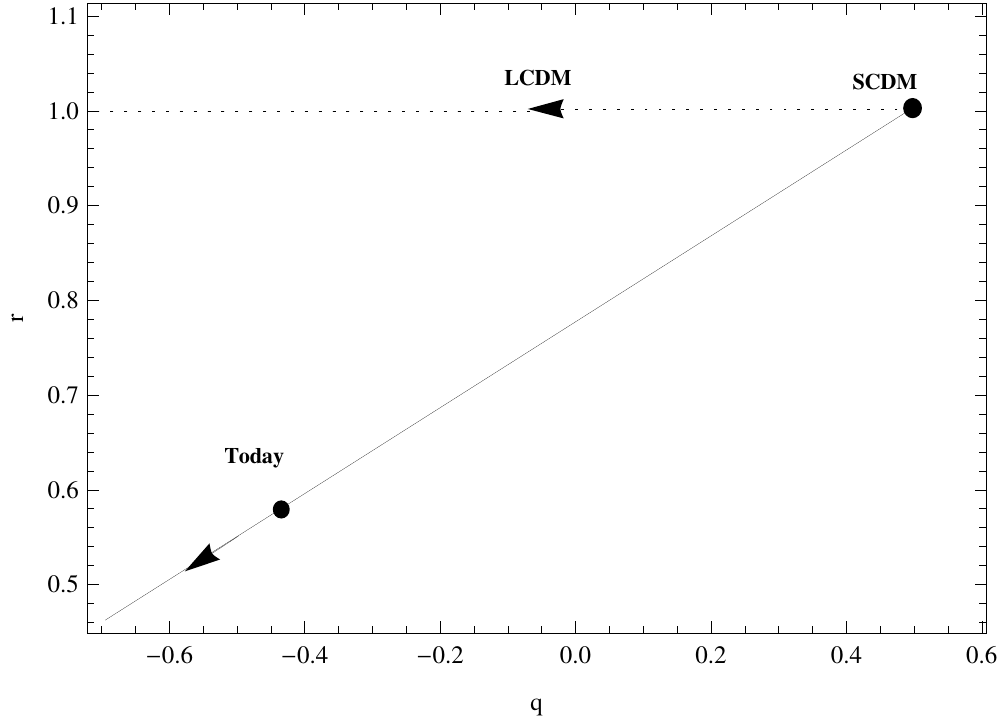}
 \caption{\emph{Evolutionary trajectory in the statefinder $r-q$ plane with 
 $\alpha=1.01$ and
 $\beta=0.15.$ The solid line represents the MHRDE model, and the dashed line the 
 $\Lambda$CDM (denoted as LCDM model in the plot) as comparison. Location of today's point is (0.59, -0.45).}}
 \label{fig:rq}
\end{figure}

\begin{figure}
 \includegraphics[scale=0.8]{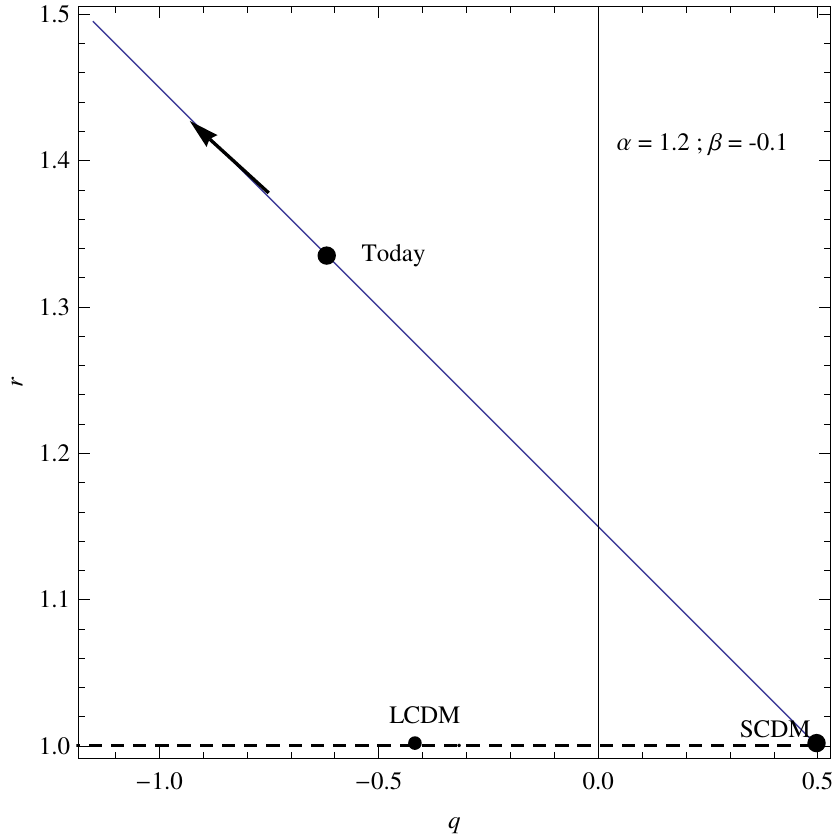}
 \caption{\emph{Evolutionary trajectory in the $r-q$ plane with $\alpha$=1.2, $\beta$=
 -0.10. The present position is denoted. The dashed shows the evolution of $\Lambda$CDM 
 (denoted as LCDM model in the plot)
 model from right to left.}}
 \label{fig:rq2}
\end{figure}

We have studied the evolutionary behavior in the $r-q$ plane also. For positive value of 
$\beta$ the plot is shown in figure \ref{fig:rq} for the 
standard values $\alpha=1.01, \, \beta=0.15$ . The figure 
shows that both $\Lambda$CDM model and MHRDE model commence evolving from the same
point in the past corresponds to $r=1, \, q=0.5,$ which corresponds to a matter 
dominated SCDM universe. In $\Lambda$CDM model the trajectory will end their evolution
at $q=-1, r=1$ which corresponds to de Sitter model, while in MHRDE model the behavior 
is different from this. 
The statefinder trajectory in holographic dark energy model with future event 
horizon has
the same starting point and the same end point as $\Lambda$CDM model 
\cite{Zimdahl1, Zhang1}.
Thus MHRDE model is
also different form holographic dark energy with event horizon form the statefinder
viewpoint.

For negative values $\beta$ the plot is as given in figure \ref{fig:rq2}. The 
evolution of the trajectory is starting from left to right. Note that the $r$ value is at the 
increase from one as the universe evolves. It is evident 
from 
the plot that the present position of the model corresponds to $r_0$=1.325 and $q_0$ =
-0.63.

\section{Conclusions}
  We have studied the modified holographic Ricci dark energy (MHRDE) in flat universe,
  where 
  the IR cutoff is given by the modified Ricci scalar, and the dark energy become
  $\rho_x = 2(\dot{H}+3\alpha H^2/2)/(\alpha - \beta)$ where $\alpha$ and $\beta$ are 
  model parameters.
  We have calculated the relevant cosmological parameters and their evolution and 
  also analyzed the model form the statefinder view point for discriminating it from 
  other models. The importance of the model is that it depends on the local quantities 
  and thus avoids the causality problem. 
  
  The density of MHRDE is comparable with the non-relativistic matter at high redshift 
  as shown in figure \ref{fig:densityevolve} and
  began to dominate at low redshifts, thus the model is free from the 
  coincidence problem.
  
  The evolution of equation of state parameter is studied.
  The equation of state parameter is nearly zero at high redshift, implies that in 
  the past 
  universe MHRDE behaves like cold dark matter. Further evolution of equation of 
  state is
  strongly depending on the model parameter $\beta.$ If the $\beta$ parameter is 
  positive
  the equation of state is greater than -1. For negative values of $\beta$, the 
  equation
  of state cross the phantom divide $\omega_x < -1.$
  
  In this model the deceleration parameter starts form around 0.5 at the early 
  times and 
  and starts to become negative when the redshift $z<1.$ . In general  
  we have found that in MHRDE model the universe entering
  the accelerating phase at times earlier (for allowed range of parameters $\alpha$
   and $\beta$), than in the $\Lambda$CDM model. But in particular as the model 
   parameter $\alpha$ 
  increases,
  the universe enter the accelerating phase at relatively later times.
  
  We have applied the statefinder diagnostic to the MHRDE and plot the trajectories 
  in the $r-s$ and $r-q$ planes. The statefinder diagnostic is a crucial tool for 
  discriminating different dark energy models. The statefinder trajectories are 
  depending on the model parameters $\alpha$ and $\beta.$ For positive values of $\beta$
  the $r$ values will decreases from one and for negative $\beta$ the $r$ will increases form
  one as the universe evolves. The values of $\alpha$ and $\beta$ are constrained using
  observational data in reference \cite{Chimento1}, the best fit value is $\alpha$=1.01, 
  $\beta$=0.15. The present value of ($r,s$) can be viewed as a discriminator for
  testing 
  different dark energy models. For the $\Lambda$CDM model statefinder is a fixed 
  point $r$=1, 
  $s$=0.  For positive values $\beta$ parameter the $r-s$ and $r-q$ plots of MHRDE 
  shows that, the evolutionary trajectories 
  starts form $r=1$ and $q=0.5,$ in the past universe (for the best fit model 
  parameters), which reveals that the MHRDE is behaving like cold dark
  matter in the past. The further evolution of MHRDE in the $r-s$ plane shows that the 
  present position of MHRDE model in the $r-s$ plane for the best fit parameter
  is $r_0$=0.59, $s_0$=0.15 and in the $r-q$ plane is $r_0$=0.59, $q_0$=-0.45. The 
  difference
  between the MHRDE and $\Lambda$CDM models is in the evolution of the equation of 
  state parameter, which is -1 in
  the $\Lambda$CDM model and a time-dependent variable in MHRDE model. A further 
  comparison can be
  made with the new HDE model \cite{Setare1}, which gives the present values 
  $r_0 (HDE)=1.357,
  s_0(HDE)=-0.102$ and $r_0(HDE)=1.357, q_0(HDE)=-0.590.$ So in the $r-s$ plane the 
  distance of the MHRDE 
  model form the $\Lambda$CDM fixed point is slightly larger compared to the new HDE 
  model for positive values of $\beta$ parameter. However in the case of MHRDE model 
  the starting point in $r-s$ plane and $r-q$ plane is 
  ($r=1, s=0$ and $r=1, q=0.5$) is same as that in the $\Lambda$CDM model. 
  
  For negative values of the $\beta$ the $r-s$ trajectory we have plotted is different 
  compared to that of positive $\beta$ values. For negative $\beta$ values the $r$ value can 
  attains values greater than one as $s$ increases. The present status of the evolution
  in 
  the $r-s$ plane is $r_0$ =1.325, $s_0$=-0.10 for model parameters $\alpha$=1.2, 
  $\beta$=-0.10 and  $r_0$=1.321, $s_0$=-0.10. The $r-q$ for $\alpha$=1.2 and 
  $\beta$=-0.10 shows that the present state 
  of the MHRDE model is corresponds to $r_0=1.325$ and $q_0$= - 0.63.
  These values shows that the MHRDE model is different form $\Lambda$CDM 
  model for the present time when $\beta$ parameter is negative also. But compared 
  the new HDE model, the present MHRDE model doesn't show much deviation, shows 
  that for negative values the behavior of MHRDE model is almost similar to new 
  HDE model. Irrespective of whether $\beta$ is positive or negative the MHRDE model is 
  commence to evolve from SCDM model. When $\beta$ is positive 1 is the maximum 
  value of $r$, on the other hand when $\beta$ is negative 1 is the minimum value of 
  $r.$ However the exact 
  discrimination of the dark energy models is possible only if we can obtain the 
  present $r-s$ values in a model independent way form the observational data. It 
  is expected 
  that the future high-precision SNAP-type observations can lead to the present 
  statefinder
  parameters, which could help us to find the right dark energy models.

\end{document}